\documentclass[twocolumn, aps, pra, 10pt]{revtex4-2}

\usepackage{graphicx}

\usepackage{amsmath, amssymb, amsfonts}

\usepackage{hyperref}
\usepackage{color}

\begin{document}

    \title{Quantum transduction of superconducting qubit in electro-optomechanical and electro-optomagnonical system.}
    \author{Roson Nongthombam}
        \email{n.roson@iitg.ac.in}
        \affiliation{Department of Physics, Indian Institute of Technology Guwahati, Guwahati-781039, India}
    \author{Pooja Kumari Gupta}
        \email{pooja.kumari@iitg.ac.in}
        \affiliation{Department of Physics, Indian Institute of Technology Guwahati, Guwahati-781039, India}
    \author{Amarendra K. Sarma}
        \email{aksarma@iitg.ac.in}
        \affiliation{Department of Physics, Indian Institute of Technology Guwahati, Guwahati-781039, India}


    \begin{abstract}
    We study the quantum transduction of a superconducting qubit to an optical photon in electro-optomechanical and electro-optomagnonical systems. The electro-optomechanical system comprises a flux-tunable transmon qubit coupled to a suspended mechanical beam, which then couples to an optical cavity. Similarly, in an electro-optomagnonical system, a flux-tunable transmon qubit is coupled to an optical whispering gallery mode via a magnon excitation in a YIG ferromagnetic sphere. In both systems, the transduction process is done in sequence. In the first sequence, the qubit states are encoded in coherent excitations of phonon/magnon modes through the phonon/magnon-qubit interaction, which is non-demolition in the qubit part. We then measure the phonon/magnon excitations, which reveal the qubit states, by counting the average number of photons in the optical cavities. The measurement of the phonon/magnon excitations can be performed at a regular intervals of time.
    \end{abstract}

    \maketitle


    \section{Introduction}
        \label{sec:intro}
        `Quantum network' is a rapidly developing area owing to its potential applications in scaling up quantum computers by connecting multiple quantum processors. Recently, much research has been initiated on developing a modular quantum computer based on linking multiple superconducting chips where each chip has a few high-quality qubits. Instead of cramping more qubits onto a single chip, which will result in high error rates and complex hardware, creating a network of modules containing few high-quality qubits on a single chip is better. This modular quantum computing approach has lower error rates and lesser hardware constraints. 
        
         For connecting the modules, optical fibers, which have low propagation loss in a noisy thermal environment, are employed. So, the qubit operations must first be transferred to the flying optical photons in the optical fiber. The transduction of the qubit to the optical photon cannot be achieved directly due to the vast separation of the frequencies between the two (qubit in GHz and optical photon in THz). One way to achieve transduction is by introducing a bosonic system as a mediator that couples both the qubit and the optical photon, forming a hybrid qubit-boson-optical system. In this work, we discuss transduction in two such hybrid systems, namely, electro-optomechanical and electro-optomagnonic systems. The electro-optomechanical system consists of a superconducting microwave circuit coupled to a mechanical resonator which in turn is connected to an optical cavity. In recent years, this hybrid system has been extensively studied experimentally\cite{Nat:Commun.11.1166,Nat.Phys.16.69,Nature.588.599,npj:Quantum:Inf.8.149,Nature:Phys.10.321} and theoretically\cite{Phys.Rev.Applied.16.064044,Phys.Rev.Applied.18.054061,Phys.Rev.A.84.042342,Phys.Rev.Lett.127.040503,Phys.Rev.A.94.012340,Phys.Rev.A.104.023509,Nat.Phys.16.257} for microwave-to-optical photon transduction.  There are several ways of coupling a transmon qubit, formed by a superconducting microwave circuit, to a mechanical resonator \cite{Nat.Commun.7.12396,Science.358.199,Science.364.368,Nature.494.211,Nature.475.359}. Here, we consider a flux tunable transmon qubit that is coupled to a suspended mechanical beam \cite{Phys.Rev.Research.2.023335}. The mechanical beam is then integrated as an end mirror of an optomechanical cavity forming the required hybrid system\cite{Phys.Rev.A.94.012340}. 

        The hybrid electro-optomagnonic system consists of a superconducting microwave circuit coupled to a ferromagnetic magnon excitation \cite{Science.349.405,Science.367.425,Phys.Rev.Lett.125.117701,YUAN20221}, which is coupled to an optical photon \cite{Phys.Rev.B.93.174427, Phys.Rev.Lett.120.133602,Phy.Rev.Lett.117.133602,Phys.Rev.Lett.129.037205,Phys.Rev.Lett.116.223601}. This hybrid system is less explored. It is mainly due to the weak coupling between the magnon and the optical photon. However, there has been some progress recently \cite{App.Phys.Express.12.070101}. For example, enhancement of magnon-photon coupling under the triple resonance condition of input photon, magnon and output photon is demonstrated in \cite{Phys.Rev.Lett.117.123605,Phy.Rev.Lett.117.133602}. By implementing the triple resonant condition, a microwave-to-optical conversion based on multiple magnon mode interaction with the optical photon mode is demonstrated in \cite{Optica.7.1291}. Another theoretical study to improve the magnon-optical coupling based on optical optical whispering gallery mode (WGM) coupled to localized vortex magnon mode in a magnetic microdisk is done in \cite{Phys.Rev.B.98.241406}.
        
        In this work, we construct the hybrid electro-optomagnonical system by merging the scheme proposed in \cite{Phys.Rev.Lett.129.037205}, where a flux-tunable transmon qubit is coupled to a magnon mode formed in a $\mu$m size YIG sphere, and the optomagnonical setup experimentally demonstrated in \cite{Phys.Rev.Lett.117.123605}, where an optical WGM interacts with magnon mode in a YIG sphere of radius having few hundred $\mu$m. One main difficulty in realizing this hybrid system is the size gap in the YIG spheres. However, the possibility of reducing the sphere used in the optomagnonic case is pointed out in \cite{Phys.Rev.Lett.117.123605}. Assuming this to be possible, we consider a YIG sphere of few $\mu$m size radius that couples both the superconducting qubit and the optical WGM present in the sphere.    

        The technique used here for measuring the qubit states from the optical photon is similar to the one employed in \cite{Nature.588.599}. The idea is to first associate or encodes the qubit states in the magnon/phonon coherent excitations and then measure these excitations by counting the average number of the photon in the optical cavity. Although measuring the qubit states by detecting the optical photon count is demonstrated in \cite{Nature.588.599}, our scheme exhibits two distinct features: (1) The interaction of the qubit and the magnon/phonon commutes with the intrinsic Hamiltonian of the qubit. In other words, the initial state of the qubit remains the same during the interaction. (2) Due to the coherent and oscillatory evolution of the magnon/phonon and photon states during the interaction, we can perform measurements of the qubit states from the photon count at regular intervals of time.   

        The paper is organized as follows. 
        We describe the two hybrid systems under study in Sec.\ref{sec:system}. The first sequence of the transduction process, i.e., encoding the qubit state in the magnon/phonon coherent excitation is studied in Sec.\ref{sec:qub_pho/mag_int}. The measurement of the qubit state from the optical photon count is done in Sec.\ref{sec:Trans2}.
        Finally, we conclude by summarizing our work in Sec. \ref{sec:conc}.


    \section{The Hybrid System}
        \label{sec:system}
\begin{figure}[t]
            \centering
            \includegraphics[width=0.4\textwidth]{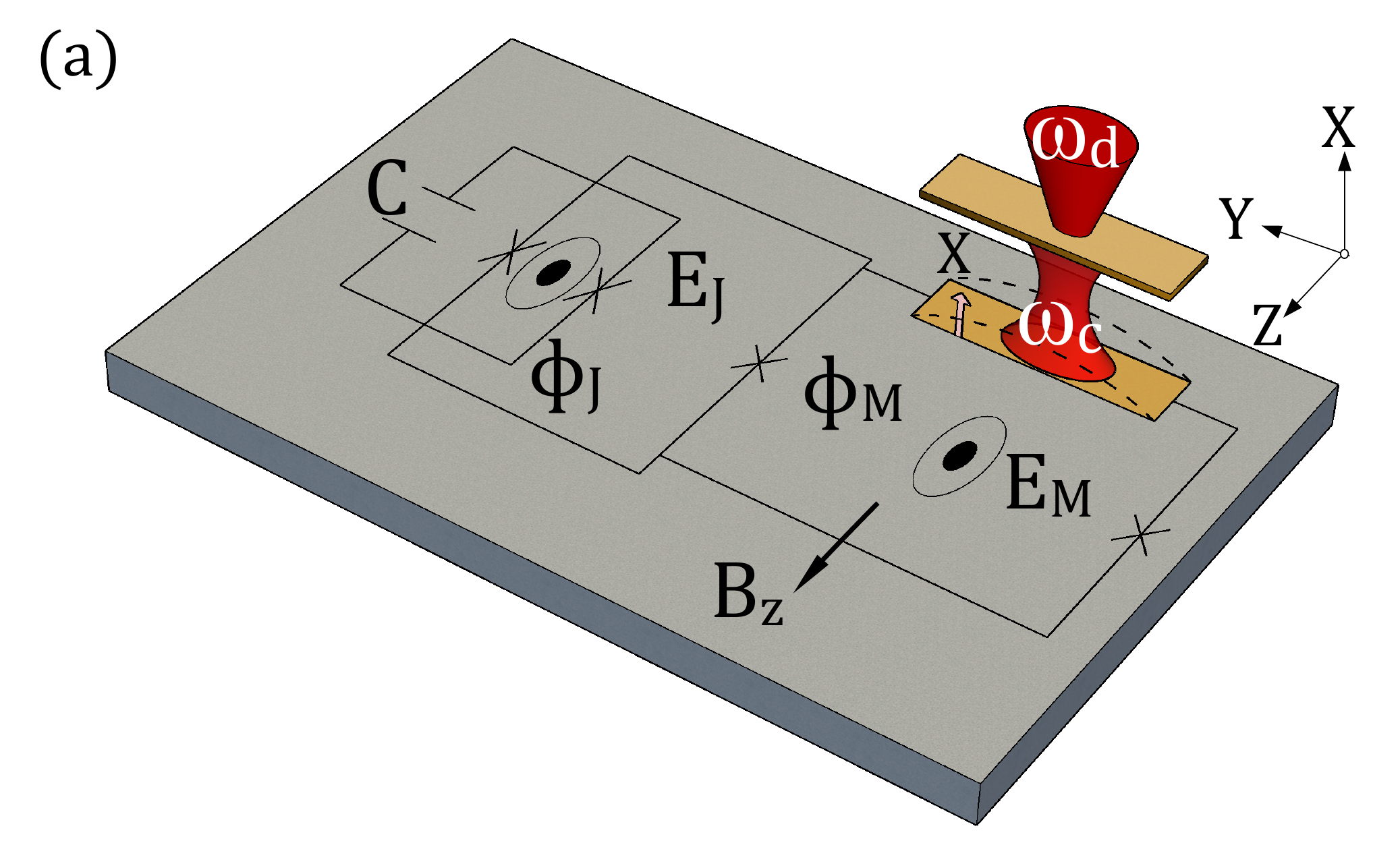}
            \includegraphics[width=0.4\textwidth]{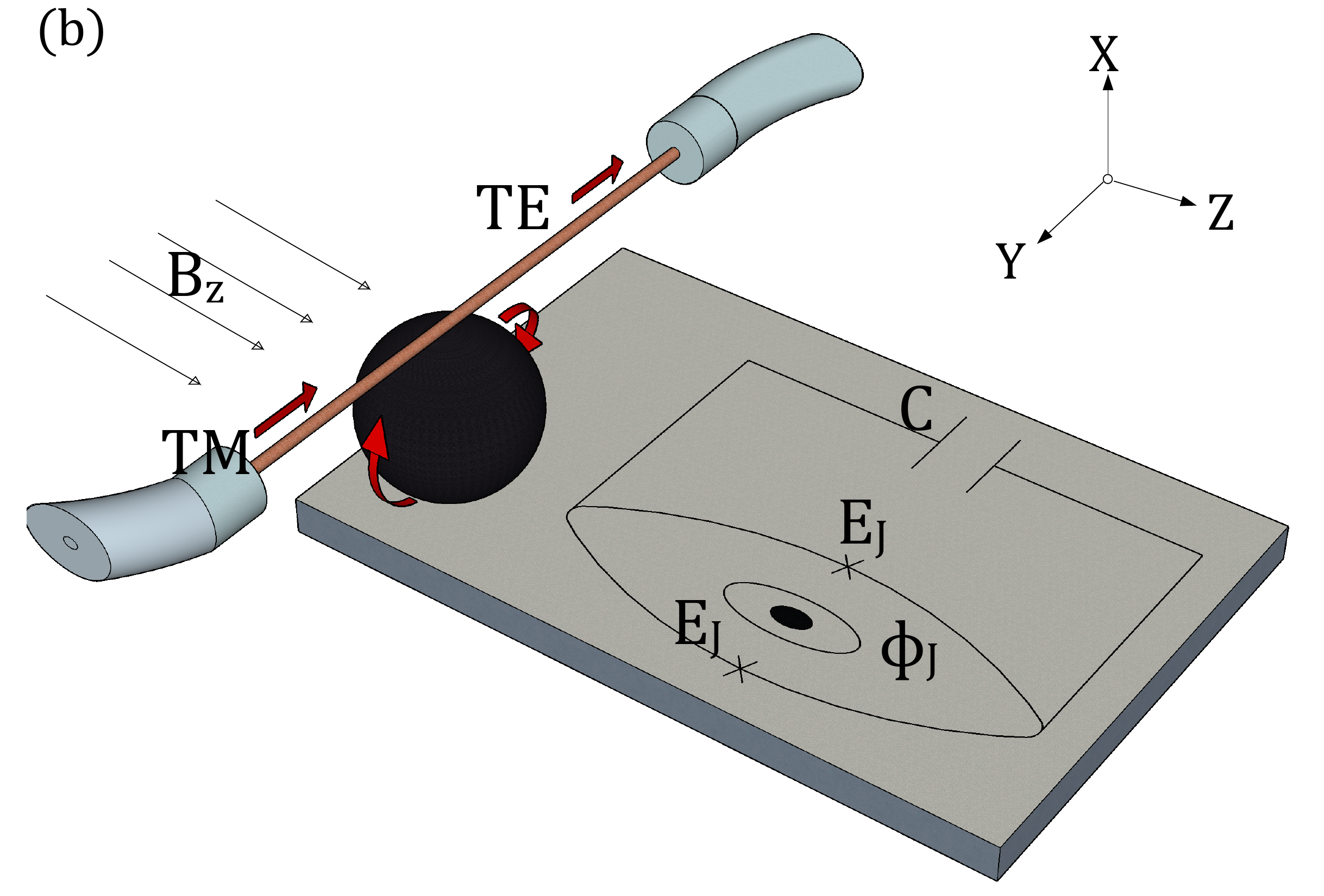}

            \caption{(Color online) (a) Schematic diagram of an electro-optomechanical system. On application of an in-plane magnetic field $B_z$, the transmon qubit formed by a SQUID loop $(E_J, \Phi_J)$ is coupled to a mechanical beam suspended at one arm of the loop $(E_M, \Phi_M)$ \cite{Phys.Rev.Research.2.023335}. The mechanical resonator is integrated as a movable plate of an optomechanical cavity whose resonance frequency is $\omega_c$. The optomechanical cavity is shined on by a red-detuned laser light. (b) A YIG ferromagnetic sphere that supports both the magnon excitation and optical WGM is placed near a flux-tunable transmon qubit formed by the loop $(E_J, \Phi_J)$ \cite{Phys.Rev.Lett.129.037205}. An optical channel is mounted on top of the YIG sphere. A TM-polarized light given at the input of the channel passes the YIG sphere in a clockwise direction, and a TE-polarized light emerges at the channel output. An in-plane magnetic field $B_z$, responsible for the transmon-magnon coupling, is passed through the ferromagnetic sphere.} 
            \label{fig:hybrid sys}
        \end{figure}

       We first consider the hybrid electro-optomechanical system. 
       This hybrid system comprises a flux-tunable transmon (formed by a SQUID loop ($E_J$, $\Phi_J$)), coupled to a mechanical resonator (realized by suspending one arm of another SQUID loop ($E_{M}$, $\Phi_{M}$))\cite{Phys.Rev.Research.2.023335} which can oscillate out of plane. The suspended mechanical membrane is then integrated as an end 
       mirror of an optical cavity forming an optomechanical cavity \cite{Phys.Rev.A.94.012340}, as shown in Fig. \ref{fig:hybrid sys}(a). The Hamiltonian of the system is described by
       \begin{equation}
           \label{eqn:ham_eom}
           \hat{H}_{eom}=\hat{H}_0+\hat{H}_{tm}+\hat{H}_{om}+\hat{H}_d
       \end{equation}
       where,
       \begin{subequations}
            \label{eqn:ham_eom2}
            \begin{eqnarray}
                \label{eqn:ham_0}
                \hat{H}_0 &=& \hbar\Delta_c\hat{a}^{\dagger}\hat{a}+\hbar\omega_m\hat{b}^{\dagger}\hat{b}+\hbar\omega_t\hat{c}^{\dagger}\hat{c}-\frac{E_c}{2}\hat{c}^{\dagger}\hat{c}^{\dagger}\hat{c}\hat{c},\\
                \label{eqn:ham_tm}
                \hat{H}_{tm} &=& \hbar g_{tm}\hat{c}^{\dagger}\hat{c}(\hat{b}+\hat{b}^{\dagger}), \\
                \label{eqn:ham_om}
                \hat{H}_{om} &=& \hbar g_{om}\hat{a}^{\dagger}\hat{a}(\hat{b}+\hat{b}^{\dagger}),\\
                \label{eqm:ham_d}
                \hat{H}_d&=&\hbar E_0 \left(\hat{a}+\hat{a}^{\dagger}\right).
            \end{eqnarray}
        \end{subequations}
        Here, $\hat{a}$($\hat{a}^{\dagger}$), $\hat{b}$($\hat{b}^{\dagger}$), and $\hat{c}$($\hat{c}^{\dagger}$)
       are the annihilation (creation) operators of the optical photon, the mechanical phonon and the transmon, respectively. $\hat{H}_0$ is the Hamiltonian of the individual components of the hybrid system in the absence of any interactions. The transmon-mechanical resonator interaction is described by $\hat{H}_{tm}$, and the optomechanical interaction by $\hat{H}_{om}$. The strength of the coupling constant $g_{om}$ is generally quite small ($\approx 1$Hz). So, we drive the optomechanical cavity to increase the coupling strength. This drive is included in the Hamiltonian of the hybrid system as $\hat{H}_d$.
       The Hamiltonian $\hat{H}_{om}$ is written in the optomechanical drive frame ($\Delta_c=\omega_c-\omega_d$, where $\omega_c$ is the cavity frequency and $\omega_d$ is the drive frequency.).
       The qubit-mechanical and optomechanical interactions arise from the displacement of the mechanical resonator. On application of an in-plane magnetic field $B_z$, as shown in Fig. 1(a), the displacement of the mechanical resonator picks up a flux in the Josephson energy. This motional dependent Josephson energy leads to the transmon-mechanical resonator interaction $\hat{H}_{tm}$ \cite{Phys.Rev.Research.2.023335}. Note that in addition to the third order non-linear interaction term, $\hat{H}_{tm}$, higher order non-linear interaction terms are present as demonstrated in \cite{Phys.Rev.Research.2.023335}. But, here, we have excluded this higher order corrections due to their negligible contribution to the system dynamics.
       The mechanical motion of the resonator also simultaneously alters the resonator frequency of the optomechanical cavity, which results in the optomechanical interaction $\hat{H}_{om}$. We have considered a small SHM (simple harmonic motion) displacement of the resonator.

       We next consider the hybrid electro-optomagnonic system. In this system, a YIG sphere having a diameter in the $\mu$m range is placed near a flux-tunable transmon formed by a symmetric SQUID loop ($E_J$,$\Phi_J$) \cite{Phys.Rev.Lett.129.037205}. The YIG sphere is then mounted on an optical fiber placed just above the plane of the SQUID loop \cite{Phys.Rev.Lett.117.123605}, as shown in Fig. \ref{fig:hybrid sys}(b). Just like in the previous system, an in-plane magnetic field $B_z$ is applied. This field magnetizes the magnetic sphere along the z-direction. Because of this magnetization, a uniform magnetostatic mode or kittle mode is excited on the YIG sphere, whose magnetic moment produces a stray field and traverse along the transmon SQUID loop. 
       Subsequently, the stray field picks up an additional flux in the loop
       thereby changing the Josephson energy and frequency of the transmon and eventually leading to a transmon-magnon interaction. On the other hand, a TM (transverse magnetic field) polarized light is pumped at the input of the optical waveguide. When in resonance, this pumped light or signal is confined in the YIG sphere forming WGM (whispering gallery mode) in the clockwise direction. The TM input signal then interacts with the magnons in the magnetic sphere. This interaction has two significant features. One is that it changes the TM-polarized input signal to a circulating TE WGM and then comes out as a TE-polarized signal at the output. The other feature is that it changes the input and output signal frequencies. The amount of change in the input and output signal frequencies is equivalent to that of the magnon frequency. The above two features are the outcomes of satisfying the triple-resonance condition. The triple resonance condition is experimentally demonstrated in a 100-300$ \mu$m size ferromagnetic sphere \cite{Phys.Rev.Lett.117.123605}. In our case, the size of the sphere is considered to be around 3 $\mu$m. Although the size we have considered here is not yet experimentally realized for the triple resonance interaction, the possibility of sizing down the sphere is pointed out in \cite{Phys.Rev.Lett.117.123605}. The effective Hamiltonian of the hybrid electro-optomagnonic system is described by
       \begin{equation}
           \label{eqn:ham_eomg}
           \hat{H}'_{eom}=\hat{H}'_0+\hat{H}'_t+\hat{H}'_{tm}+\hat{H}'_{om}+\hat{H}'_d
       \end{equation}
       where,
       \begin{subequations}
            \begin{eqnarray}
                \label{eqn:ham_01}
                \hat{H}'_0 &=& \hbar\omega_v\hat{a}_v^{\dagger}\hat{a}_v+\hbar\omega_h\hat{a}_h^{\dagger}\hat{a}_h+\hbar\omega'_{m}\hat{m}^{\dagger}\hat{m},\\
                \label{eqn:ham_t1}
                \hat{H}'_t&=&\hbar\omega'_t\hat{c}^{\dagger}\hat{c}-\frac{E_c}{2}\hat{c}^{\dagger}\hat{c}^{\dagger}\hat{c}\hat{c},\\
                \label{eqn:ham_tm1}
                \hat{H}'_{om} &=& \hbar g'_{om}(\hat{a}_h^{\dagger}\hat{a}_v\hat{m}+\hat{a}_h\hat{m}^{\dagger}\hat{a}_v^{\dagger}), \\
                \label{eqn:ham_om1}
                \hat{H}'_{tm} &=& \hbar g'_{tm}\hat{c}^{\dagger}\hat{c}(\hat{m}+\hat{m}^{\dagger}),\\
                \label{eqm:ham_d1}
                \hat{H}'_d&=&\hbar E_v \left(\hat{a}_ve^{-i\omega_Lt}+\hat{a}_v^{\dagger}e^{i\omega_Lt}\right).
            \end{eqnarray}
        \end{subequations}
        Here, $\hat{a}_v$($\hat{a}_v^{\dagger}$), $\hat{a}_h$($\hat{a}_h^{\dagger}$), $\hat{m}$($\hat{m}^{\dagger}$), and $\hat{c}$($\hat{c}^{\dagger}$)
       are the annihilation (creation) operators of the input TM optical photon, the output TE optical photon, the magnon and the transmon, respectively. $\hat{H}'_0$ and $\hat{H}'_t$ are the Hamiltonian of the individual components of the hybrid system in the absence of any interactions. The transmon-magnon interaction is described by $\hat{H}'_{tm}$ and the  optomagnonic interaction by $\hat{H}'_{om}$.
       Here, the interaction term $\hat{H}'_{tm}$ is for the symmetric SQUID loop. $\hat{H}'_d$ is the optical drive of the TM mode.

    \section{Transduction}
        \label{sec:Trans}
       Here, we discuss the quantum transduction of qubit states to optical photons via mechanical phonons or YIG sphere magnons. The transduction process is realized in sequence. First, we encode the qubit states to the phonon/magnon excitations, and next, we measure these excitations by counting the average number of photons in the optical cavity. This section is divided into two parts. The first part discusses the process of encoding the qubit states in the mechanical phonon states, and in the second part, we discuss how the phonon states, and hence the qubit states, are determined from the optical photon number.
       
       \subsection{Qubit-phonon/magnon transduction}
        \label{sec:qub_pho/mag_int}
             We first consider the qubit-mechanical interaction in the hybrid electro-optomechanical system and show how qubit states can be encoded to the phonon excitations. The qubit-mechanical coupling rate $g_{tm}$ is much larger than the single-photon optomechanical coupling rate $g_{om}$. So, if there is no optomechanical cavity drive, we can neglect the optomechanical interaction in the Hamiltonian given by Eq. \ref{eqn:ham_eom2}. Now we are left with just the electromechanical part of the hybrid system.
            \begin{equation}
            \label{eqn:ham_em}
                \hat{H}_{em}=\hat{H}_0+\hbar g_{tm}\hat{c}^{\dagger}\hat{c}(\hat{b}+\hat{b}^{\dagger}).
            \end{equation}
            Here, the coupling constant $g_{tm}$ is dependent on the external flux bias $\Phi_m$ as \cite{Phys.Rev.Lett.129.037205}, 
            \begin{equation}
            \label{eqn:g_tm}
                g_{tm}=g_0 sin(\phi_b),
            \end{equation}
            \\
             where, $\phi_b=\frac{\pi\Phi_b}{\Phi_0}$ and $\Phi_0=\frac{h}{2e}$ is the flux quantum. $g_0$ is coupling constant. Next, we enhance the coupling rate by modulating it parametrically by applying a weak ac bias $\phi_b=\phi_{ac}cos(\omega_{ac}t)$ $(\phi_{ac}<<1)$ as done in \cite{Phys.Rev.Lett.129.037205}.
            \begin{equation}
             \label{eqn:g_tm2}
                g_{tm}=g_0\phi_{ac}cos(\omega_{ac}t).
            \end{equation}
            By substituting this modulated time dependent coupling constant in the Hamiltonian \ref{eqn:ham_em}, and then transforming the resultant Hamiltonian in the reference frame of the ac drive $(U=e^{i\omega_{ac}tb^{\dagger}b })$, we get
            \begin{equation}
            \label{eqn:ham_em2}
                \hat{H}'_{em}=\hat{H}_0+\hbar g_0\phi_{ac}\hat{c}^{\dagger}\hat{c}(\hat{b}+\hat{b}^{\dagger})-\omega_{ac}\hat{b}^{\dagger}\hat{b}.
            \end{equation}
            Here, we have ignored the fast rotating terms since $2\omega_{ac}>>g_0\phi_{ac}$. To do the qubit transduction, we convert the transmon to a transmon qubit by considering only the first two energy levels. We then let the system evolve under resonant modulation $(\omega_m=\omega_{ac})$. If the qubit is initially in the ground state $|g\rangle$ and the mechanical resonator is in the vacuum state $|0_b\rangle$ then after some time t, the qubit will remain in the ground state and the mechanical resonator will change to a coherent state $|\beta_b=ig_0\phi_{ac}t\rangle$. Similarly, if the qubit is initially in the excited state $|e\rangle$ and the resonator in the vacuum state, then the qubit will remain in the excited state, and the mechanical resonator will evolve to another coherent state $|\beta_b=-ig_0\phi_{ac}t\rangle$ after some time t.
            \begin{center}
                $|g,0_b\rangle_0\longrightarrow|g,\beta_b=g_0\phi_{ac}t\rangle_t$\\
                \vspace{0.4cm}
                $|e,0_b\rangle_0\longrightarrow|e,\beta_b=-ig_0\phi_{ac}t\rangle_t$
            \end{center}
            An overall phase term induced from the intrinsic qubit Hamiltonian is not included as it does not contribute to the transduction process. We see that as the system evolves, the mechanical resonator changes from a vacuum state to a coherent state, whereas the qubit state remains as it is. It is because the interaction between the qubit and the mechanical resonator commute with the intrinsic Hamiltonian of the qubit. In other words, the interaction is `non-demolition' in the qubit part.
            
            We next consider transduction in the electro-optomagnonic case and analyze how qubit states can be encoded to magnon excitations. Just like in the previous case, we can neglect the optomagnonic part and consider only the electro-magnonic part since the single magnon-photon coupling $g'_{om}$ is much less than the transmon-magnon coupling $g'_{tm}$ for no optical drive. The electro-magnonic part is described by
            \begin{equation}
            \label{eqn:ham'_em}
                \hat{H}'_{em}=\hat{H}'_0+\hbar g'_{tm}\hat{c}^{\dagger}\hat{c}(\hat{m}+\hat{m}^{\dagger}),
            \end{equation}
            where, 
            \begin{equation}
            \label{eqn:g'_tm}
                g'_{tm}=g'_0 \frac{sin(\phi_m)}{\sqrt{|cos(\phi_m)|}},
            \end{equation}
            where, $\phi_m=\frac{\pi\Phi_m}{\Phi_0}$ and $\Phi_0=\frac{h}{2e}$ is the flux quantum. $g_0$ is coupling constant. Similar to the previous system, here also we enhance the coupling rate by modulating it parametrically by applying a weak ac bias $\phi_m=\phi'_{ac}cos(\omega'_{ac}t)$ $(\phi'_{ac}<<1)$ as done in \cite{Phys.Rev.Lett.129.037205}.
            \begin{equation}
             \label{eqn:g'_tm2}
                g'_{tm}=g'_0\phi'_{ac}cos(\omega'_{ac}t).
            \end{equation}
            Substituting Eq. \ref{eqn:g'_tm2} in Eq. \ref{eqn:ham'_em} and then transforming in drive frame $(U=e^{i\omega'_{ac}tm^{\dagger}m })$ gives
            \begin{equation}
            \label{eqn:ham_em2}
                \hat{H}'_{em}=\hat{H}_0+\hbar g'_0\phi'_{ac}\hat{c}^{\dagger}\hat{c}(\hat{m}+\hat{m}^{\dagger})-\omega'_{ac}\hat{m}^{\dagger}\hat{m}.
            \end{equation}
            The fast rotating terms are neglected for $2\omega'_{ac}>>g'_0\phi'_{ac}$. We now take the first two levels of the transmon and allow the system to evolve. For resonance modulation $(\omega'_m=\omega'_{ac})$, we obtain results similar to that of the electro-mechanical case, i.e.,
            \begin{center}
                $|g,0_m\rangle_0\longrightarrow|g,\beta_m=ig'_0\phi'_{ac}t\rangle_t$\\
                \vspace{0.4cm}
                $|e,0_m\rangle_0\longrightarrow|e,\beta_m=-ig'_0\phi'_{ac}t\rangle_t$
            \end{center}
            Here, $|g,0_m\rangle_0$ and $|e,0_m\rangle_0$ are the initial states, and $|g,\beta_m=ig'_0\phi'_{ac}t$ and $|e,\beta_m=-ig'_0\phi'_{ac}t\rangle_t$ are the final states of the qubit-magnonic system after time t. An overall phase is not included.

\begin{figure}[t]
            \centering
            \includegraphics[width=0.23\textwidth]{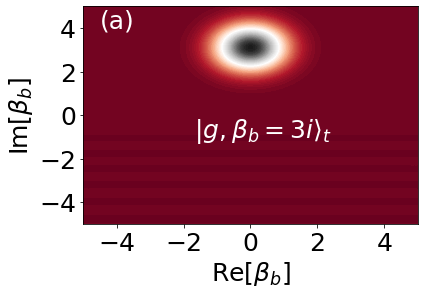}
            \includegraphics[width=0.23\textwidth]{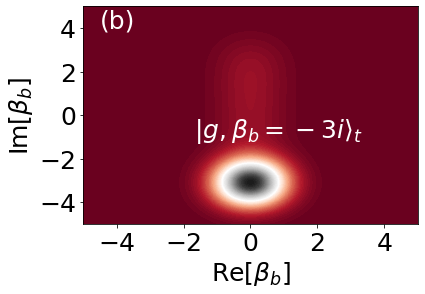}
            \includegraphics[width=0.23\textwidth]{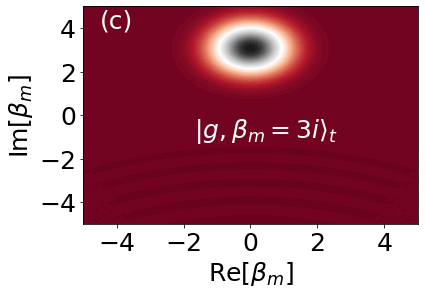}
            \includegraphics[width=0.23\textwidth]{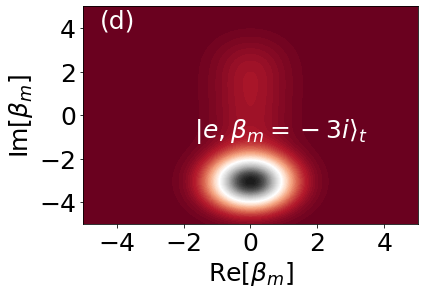}
            \caption{(Color online) Wigner function representation of the coherent states of phonon and magnon. In (a) and (b), the phonon and the magnon excites to the coherent states $|\beta_b=3i\rangle_t$ and $|\beta_m=3i\rangle_t$ when the qubit is in the ground state $|g\rangle$. The coherent states of the phonon $|\beta_b=-3i\rangle_t$ and the magnon $|\beta_m=-3i\rangle_t$ when the qubit is in the exited state $|e\rangle$ is shown in (c) and (d), respectively. The coherent states are taken at time $t=\tau=(3/2\pi) $ $\mu s$ for coupling constants $g_0\phi_{ac}=g'_0\phi'_{ac}=2\pi$ MHz. The other parameters are $gamma_b/2\pi=1$ Hz, $n_{th}=400$, $gamma_m/2\pi=0.1$ GHz, $n'_{th}=0.5$, $\Gamma/2\pi=0.1$ GHz } 
            \label{fig:mech_mag evo}
            \end{figure}

            So far, we have not included the noise factor while evolving the system. To include the noisy environment, we allow the system to evolve under the Lindblad master equation. 
            For the electro-mechanical system
            \begin{eqnarray}
                \label{eqn:master1}
                \dot{\hat{\rho}}_{em} &=& -i [ \hat{H}_{em},\hat{\rho}_{em} ] + \Gamma \mathcal{L} [ \hat{\sigma_z} ] +  \Gamma \mathcal{L} [ \hat{\sigma}^- ] \nonumber \\
                && + \gamma_b (n_{th}+1)\mathcal{L} [ \hat{b} ]+\gamma_b n_{th}\mathcal{L} [ \hat{b}^{\dagger} ],
            \end{eqnarray}
            and for the electro-magnonic system
            \begin{eqnarray}
                \label{eqn:master2}
                \dot{\hat{\rho}}'_{em} &=& -i [ \hat{H}'_{em},\hat{\rho}'_{em} ] + \Gamma \mathcal{L} [ \hat{\sigma_z} ] +  \Gamma \mathcal{L} [ \hat{\sigma}^- ] \nonumber \\
                && + \gamma_m (n'_{th}+1)\mathcal{L} [ \hat{m} ]+\gamma_m n'_{th}\mathcal{L} [ \hat{m}^{\dagger} ],
            \end{eqnarray}
            where $\mathcal{L} [ \hat{o} ] = ( 2 \hat{o} \hat{\rho} \hat{o}^{\dagger} - \hat{o}^{\dagger} \hat{o} \hat{\rho} - \hat{\rho} \hat{o}^{\dagger} \hat{o} ) / 2$. Here, $\Gamma$ is the decay rate of the transmon qubit, $\gamma_b$($\gamma_m$) is the decay rate of phonon (magnon), $n_{th}$($n'_{th}$) is the thermal phonon (magnon) number, and $\hat{\rho}_{em}$($\hat{\rho}'_{em}$) is the density operator of the qubit-mechanical (qubit-magnonic) system. To observe the coherent excitations of phonon and magnon in the dissipating environment, we plot the Wigner functions in Fig \ref{fig:mech_mag evo}. Here, we observe that the Wigner functions of the phonon and magnon at some time $\tau=(3/2\pi)$ $\mu s$ and for coupling constants $g_0\phi_{ac}=g'_0\phi'_{ac}=2\pi$ MHz show coherent state profile. The amplitude of the coherent states when the qubit is in the ground state is $|\beta_b=ig_0\phi_{ac}\tau\rangle=|3i\rangle$ for the phonon and $|\beta_m=g'_0\phi'_{ac}\tau\rangle=|3i\rangle$ for the magnon, as shown in the figure. On the other hand, when the qubit is in the excited state, the coherent amplitudes are $|\beta_b=-ig_0\phi_{ac}\tau\rangle=|3i\rangle$ and $|\beta_m=-ig'_0\phi'_{ac}\tau\rangle=|3i\rangle$. These changes in the amplitude of the coherent states corresponding to the qubit ground and excited states are similar to the ones that are observed in the non-dissipative case. 
             
            So, in both the dissipative and non-dissipative qubit-mechanical and qubit-magnonic systems, we observe that the ground state of the qubit is encoded or associated with a coherent excitation of both the phonon and magnon and the excited state of the qubit is encoded in another coherent excitation of the same magnon and phonon having amplitudes exactly opposite to that of the excitation associated with qubit ground state.
            
        \subsection{Qubit-optical photon transduction}
        \label{sec:Trans2}
        We have seen that the state of the qubit can be encoded in the coherent excitations of phonon and magnon. Here, we will complete the qubit transduction sequence by transferring the phonon and magnon states to the optical photon. In the phonon case, this can be achieved through the optomechanical interaction, and in the magnon case, it can be achieved through the optomagnonic interaction satisfying the triple-resonant condition. 

        First, we consider the optomechanical transfer. We have previously seen from the electro-mechanical interaction that the mechanical resonator can be coherently excited with different amplitudes depending on the initial states of the qubit. So, we first excite the mechanical resonator to coherent states $|\beta_b=\pm ig_0\phi_{ac}t\rangle$ and then switch off the interaction $g_0\phi_{ac}$ by turning off the flux bias $\phi_{ac}$.
        The interacting system remaining is then the optomechanical system.
        \begin{equation}
        \label{eqn:ham_op1}
            \hat{H}=\hbar\Delta_c\hat{a}^{\dagger}\hat{a}+\hbar\omega_m\hat{b}^{\dagger}\hat{b}+\hbar g_{om}\hat{a}^{\dagger}\hat{a}(\hat{b}^{\dagger}+\hat{b}) +\hbar E_0(\hat{a}^{\dagger}+\hat{a}).
        \end{equation}
        Since $g_{om}\approx1$Hz is very weak, we drive the cavity with an intense laser. Because of this strong drive, we can separate the amplitudes of the mechanical resonator and optical cavity into a semi-classical coherent part $(\beta, \alpha)$ and a small quantum fluctuation $(\delta \hat{a}, \delta \hat{b})$ around it, i.e., $\hat{a}\rightarrow \delta\hat{a}+\alpha$ and $\hat{b}\rightarrow \delta\hat{b}+\beta$. We substitute this separation in Eq.\ref{eqn:ham_op1}. By retaining only the interacting term, which is multiplied by the factor $\alpha (|\alpha|\approx10^3)$, the Hamiltonian reads
        \begin{equation}
            \label{eqn:ham_op2}
            \hat{H}_{om}=\hbar\Delta'\hat{a}^{\dagger}\hat{a}+\hbar\omega_m\hat{b}^{\dagger}\hat{b}+\hbar G_{om}(\hat{a}^{\dagger}+\hat{a})(\hat{b}^{\dagger}+\hat{b}),
        \end{equation}
        where $\Delta=\Delta_c-(\beta+\beta^*)g_{om}$ and $G_{om}=g_{om}|\alpha|$ for a constant phase preference of alpha. For simplicity we have rewritten $\delta\hat{a}$ to $\hat{a}$ and $\delta\hat{b}$ to $\hat{b}$. Note that while writing Eq.\ref{eqn:ham_op2}, we have ignored all the constant terms and all the linear terms containing $\hat{a}$, $\hat{a}^{\dagger}$, $\hat{b}$ and $\hat{b}^{\dagger}$ are equated to zero \cite{bowen2015quantum}. 

\begin{figure}[t]
            \centering
            \includegraphics[width=0.23\textwidth]{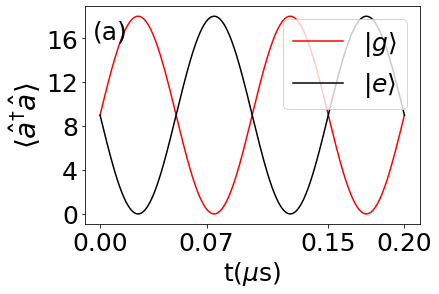}
            \includegraphics[width=0.23\textwidth]{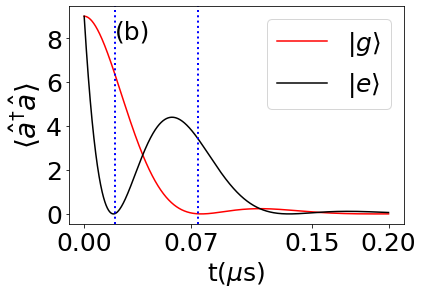}
            \includegraphics[width=0.23\textwidth]{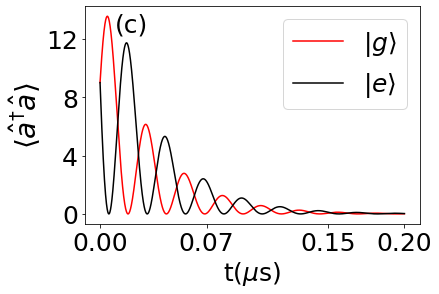}
            \includegraphics[width=0.23\textwidth]{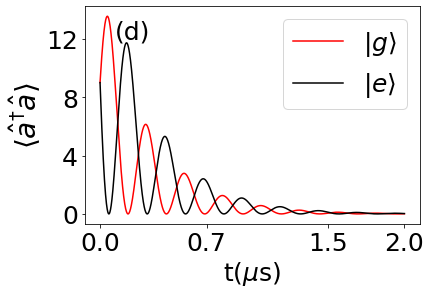}
            \caption{(Color online) Evolution of the average number of photon $\langle\hat{a}^{\dagger}\hat{a}\rangle$ in the optical cavity. (a) In the absence of dissipation, the oscillatory evolution of $\langle\hat{a}^{\dagger}\hat{a}\rangle$ keeps on going. When the qubit is in the ground state, the oscillation is represented by the red colour, and when the qubit is in the ground state, the oscillation is represented by the black colour. The evolution of average photon number in the presence of dissipation is shown in (b) for $\kappa=2G_{om}$, (c) and (d) for $\kappa=G_{om}$. In (a), (b), and (c), $\kappa/2\pi=0.01$ GHz is used, and in (d), $\kappa/2\pi=1$ MHz is used. The other common parameters are $\gamma=1$ Hz and $n_{th}=400$ }  
            \label{fig:nb evolution}
            \end{figure}

        The coherent state of the mechanical resonator prepared from the electro-mechanical interaction is in the mechanical frame $\omega_m=\omega_{ac}$. So, we transform the Hamiltonian \ref{eqn:ham_op2} in the mechanical frame. We further transform the system in the cavity detuning frame $\Delta$. Therefore, for a red-detuned laser drive $\Delta=\omega_m$, Eq.\ref{eqn:ham_op2} becomes
        \begin{equation}
            \label{eqn:ham_op3}
            \hat{H}_{om}=\hbar G_{om}(\hat{a}^{\dagger}\hat{b}+\hat{b}^{\dagger}\hat{a}).
        \end{equation}
        Here, the fast rotating terms are ignored provided $G_{om}<<2\omega_m$. For studying the state transfer from mechanical phonon to optical photon, we write down the dynamics of average number of photon and phonon in the presence of dissipation.
        \begin{subequations}
            \label{eqn:op_dyna}
            \begin{eqnarray}
                \frac{d\langle\hat{a}^{\dagger}\hat{a}\rangle}{dt}&=&-i(\langle\hat{a}^{\dagger}\hat{b}\rangle-\langle\hat{b}^{\dagger}\hat{a}\rangle)G_{om}-\kappa \langle\hat{a}^{\dagger}\hat{a}\rangle \\
                \frac{d\langle\hat{b}^{\dagger}\hat{b}\rangle}{dt}&=&-i(\langle\hat{b}^{\dagger}\hat{a}\rangle-\langle\hat{a}^{\dagger}\hat{b}\rangle)G_{om}-\gamma_b \langle\hat{b}^{\dagger}\hat{b}\rangle\nonumber\\ && +\gamma_b n_{th}\\
                \frac{d\langle\hat{b}^{\dagger}\hat{a}\rangle}{dt}&=&\frac{-(\kappa+\gamma_b)}{2}\langle\hat{b}^{\dagger}\hat{a}\rangle-iG_{om}(\langle\hat{a}^{\dagger}\hat{a}\rangle\nonumber\\&&-\langle\hat{b}^{\dagger}\hat{b}\rangle)
            \end{eqnarray}
        \end{subequations}
        By choosing the initial state of the mechanical resonator as the coherent state $|\beta_b(0)\rangle$ prepared from the electro-mechanical interaction, the average number of photon in the absence of dissipation is given by
        \begin{equation}
            \label{eqn:photon_no1}
            \langle\hat{a}^{\dagger}\hat{a}(t)\rangle=(g_{0}\phi_{ac}\tau)^2(1-sin(2G_{om}t))
        \end{equation}
        when the qubit is in the ground state ($|\beta_b(0)\rangle=|+ig_{0}\phi_{ac}\tau\rangle$), and 
        \begin{equation}
            \label{eqn:photon_no2}
            \langle\hat{a}^{\dagger}\hat{a}(t)\rangle=(g_{0}\phi_{ac}\tau)^2(1+sin(2G_{om}t))
        \end{equation}
        when the qubit is in the excited state ($|\beta_b(0)\rangle=|-ig_{0}\phi_{ac}\tau\rangle$).
        Here, we have taken the initial state of the cavity photon to be $|\alpha(0)\rangle=|g_{0}\phi_{ac}\tau\rangle$. The reason for choosing this particular initial state is discussed in the Appendix \ref{app:dynamics}. The evolution of the average photon number is shown in Fig. \ref{fig:nb evolution}(a). From the figure, we observe that if we measure the average photon number in the cavity at the interval of $t=\pi/2G_{om}$ (starting from $t=\pi/4G_{om}$) , then we either detect or do not detect the presence of photons depending on the state of the qubit. If we detect photons in the cavity at the interval of $t=(2n+1)\pi/4G_{om}$, where $n=0, 2, 4, ...$, then we know that the qubit is in the ground state, and if at the same interval, we do not detect any photons then the qubit is in the excited state. Similarly, if we detect photons in the cavity at the interval of $t=(2n+1)\pi/4G_{om}$, where $n=1, 3, 5, ...$, then we know that the qubit is in the excited state, and if at the same interval, we do not detect any photons then the qubit is in the ground state. We have chosen the above particular intervals because the average photon numbers at these intervals are at the maximum separation, and the qubit states can be determined more efficiently than the other intervals.

        In the presence of dissipation, the oscillatory nature of $\langle\hat{a}^{\dagger}\hat{a}(t)\rangle$ decays with time, and in order to know the state of the qubit by counting the photon number we require that the optomechanical coupling rate $G_{om}$ should be comparable to the decay rate $\kappa$ of the cavity. In Fig. \ref{fig:nb evolution}(b), we show the decay of cavity photon number for $\kappa=2G_{om}$, a moderate coupling strength. At this coupling strength, we are able to make an efficient measurement of qubit states at just two intervals, $t=0.02\mu s$ and $t=0.075\mu s$, before the number of average photon decay to zero. At a coupling strength lower than this, we will not be able to identify the qubit states from the optical photon. We also plot the case when the coupling strength is equal to the decay rate in Fig. \ref{fig:nb evolution}(c). Here, more oscillations can be seen, and hence more time intervals to measure the qubit states. Furthermore, we can increase the time period for a same number of oscillations by decreasing the decay rate $\kappa$ as shown in Fig. \ref{fig:nb evolution}(d). 

\begin{figure}[t]
            \centering
            \includegraphics[width=0.23\textwidth]{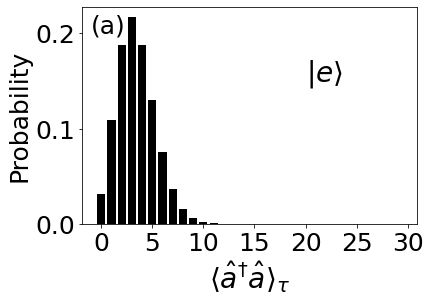}
            \includegraphics[width=0.23\textwidth]{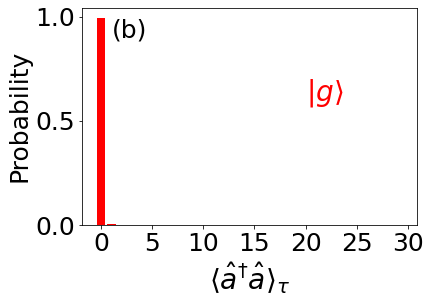}
            \caption{(Color online) Probability distribution of coherent states of the optical photon number in the presence of dissipation. (a) and (b) shows the distribution when the qubit is in the ground and excited state, respectively. The coherent states are measured at time $\tau=0.075$ $\mu$s. The average photon number in (a) is 3.4, and 0 in (b) } 
            \label{fig:nb dis}
            \end{figure}

        One could go on and find out the fidelity of state transfer of coherent state from the mechanical phonon to the optical photon. However, in our case, it is not necessary since our purpose of determining the qubit state is achieved by simply counting the cavity photon number. Since we are dealing with coherent states, we can quantify how well the measured photon number indicates that the qubit is in a particular state. In Fig. \ref{fig:nb dis}, we plot the probability distribution of the coherent state for the $\kappa=2G_{om}$ coupling case (Fig. \ref{fig:nb evolution}(b)) at the measurement time $\tau=0.075$ $\mu$s. We see that even when the qubit is in the excited state, there is still some probability of not finding any  photons in the cavity. The difference in the probability of not finding photons in the cavity when the qubit is in the excited state $(P_e=0.035)$ and when it is in the ground state $(P_g=0.999)$ gives the efficiency of determining the qubit state, $P=P_e-P_g=0.964$. This efficiency decreases for less average photon number and vice versa. So, we need to repeat the counting measurement several times before concluding the nature of the qubit state.  
            
        We now move on to the optomagnonic state transfer. Just like in the optomechanical case, we first excite the magnon to coherent state $|\beta_m=\pm ig'_{om}\phi'_{om}t\rangle$ for some time t, and then switch off the interaction $g'_{om}\phi'_{om}$ by turning off the flux bias $\phi'_{om}$. The remaining optomagnonic system in the drive frame then reads
        \begin{eqnarray}
            \label{eqn:ham_0mg1}
            \hat{H}'_{om} &=& \hbar\delta_v\hat{a}_v^{\dagger}\hat{a}_v+\hbar\delta_h\hat{a}_h^{\dagger}\hat{a}_h+\hbar\omega'_{m}\hat{m}^{\dagger}\hat{m}+\hbar E_v \left(\hat{a}_v+\hat{a}_v^{\dagger}\right) \nonumber\\
            &&\hbar g'_{om}(\hat{a}_h^{\dagger}\hat{a}_v\hat{m}+\hat{a}_h\hat{m}^{\dagger}\hat{a}_v^{\dagger}),
        \end{eqnarray}
        where $\delta_h=\omega_h-\omega_L$ and $\delta_v=\omega_v-\omega_L$. We write down the dynamics of the system.
        \begin{subequations}
            \label{eqn:opmag_dyna}
            \begin{eqnarray}
                \frac{d\hat{m}}{dt}&=&-(\frac{\gamma_m}{2}+i\omega'_m)\hat{m}+ig'_{om}\hat{a}_h\hat{a}^{\dagger}_v,\\
                \frac{d\hat{a}_v}{dt}&=&-(\frac{\kappa_v}{2}+i\delta_v)\hat{a}_v+ig'_{om}\hat{a}_h\hat{m}^{\dagger}+iE_v,\\
                 \frac{d\hat{a}_h}{dt}&=&-(\frac{\kappa_h}{2}+i\delta_h)\hat{a}_h+ig'_{om}\hat{a}_v\hat{m}.
            \end{eqnarray}
        \end{subequations}
        Here, $\gamma_m$, $\kappa_v$ and $\kappa_h$ are the decay rates of magnon, input TM field, and output TE field, respectively. The intrinsic magnon-photon coupling rate $g'_{om}$ is of the order of 10Hz, which is relatively very weak. We can enhance this coupling strength up to the order of MHz by performing an optical drive $(E_v)$ to the ferromagnetic sphere. After the drive, we can separate the input field into semi-classical mean amplitude $(\alpha_v)$ and small quantum fluctuation around it $(\delta\hat{a}_v)$, i.e.,  $\hat{a}_v\rightarrow\alpha_v+\delta\hat{a}_v$. Substituting this separation in Eq. \ref{eqn:opmag_dyna} and writing the quantum and classical parts separately, we have
        \begin{subequations}
            \label{eqn:opmag_dyna2}
            \begin{eqnarray}
                \frac{d\hat{m}}{dt}&=&-(\frac{\gamma_m}{2}+i\omega'_m)\hat{m}+ig'_{om}(\hat{a}_h\hat{a}^{\dagger}_v+\hat{a}_h\alpha^*_v),\\
                \frac{d\hat{a}_v}{dt}&=&-(\frac{\kappa_v}{2}+i\delta_v)\hat{a}_v+ig'_{om}(\hat{a}_h\hat{m}^{\dagger}+\alpha_h\hat{m}^{\dagger}),\\
                 \frac{d\hat{a}_h}{dt}&=&-(\frac{\kappa_h}{2}+i\delta_h)\hat{a}_h+ig'_{om}\hat{a}_v\hat{m},
            \end{eqnarray}
        \end{subequations}
        and 
        \begin{eqnarray}
            \label{eqn:class}
            \frac{d\alpha_v}{dt}=-(\frac{\kappa_v}{2}+i\delta_v)\alpha_v+iE_v.
        \end{eqnarray}
        The linear coupling terms in Eq. \ref{eqn:opmag_dyna2} are multiplied by a factor of $\alpha_v$ or $\alpha^*_v$ $(|\alpha_v|\approx10^3)$ compared to the non-linear coupling terms. Therefore, we can neglect the non-linear coupling terms and retain only the linear coupling terms. The corresponding linear Hamiltonian reads
        \begin{eqnarray}
            \label{eqn:ham_0mg2}
            \hat{H}'_{om} &=& \hbar\delta_v\hat{a}_v^{\dagger}\hat{a}_v+\hbar\delta_h\hat{a}_h^{\dagger}\hat{a}_h+\hbar\omega'_{m}\hat{m}^{\dagger}\hat{m}+ \nonumber\\
            &&\hbar G'_{om}(\hat{a}^{\dagger}_h\hat{m}+\hat{m}^{\dagger}\hat{a}_h),
        \end{eqnarray}
        where $G'_{om}=g'_{om}|\alpha_v|$. $|\alpha_v|$ is given by the steady value of Eq. \ref{eqn:class}. 

        Since the initial coherent state of the magnon prepared from the electro-magnonic interaction is in the magnon frame, we transform the Hamiltonian \ref{eqn:ham_0mg2} in the magnon frame. Thus, for a resonant optical drive $\delta_v=0$, the resultant Hamiltonian of the system in the magnon as well as the output TE field frame of reference yields
        \begin{eqnarray}
            \label{eqn:ham_0mg3}
            \hat{H}'_{om} =\hbar G'_{om}(\hat{a}^{\dagger}_h\hat{m}+\hat{m}^{\dagger}\hat{a}_h),
        \end{eqnarray}
        Here, since the interaction satisfies the triple resonance condition, we have taken $\delta_h=\omega'_m$ or $\omega_h=\omega_v+\omega'_m$ and ignored the fast-rotating terms provided $2\omega'_m>>G'_{om}$. We see that Hamiltonian \ref{eqn:ham_0mg3} and \ref{eqn:ham_op3} are identical. Therefore, the analysis that we have done for determining the qubit states in the optomechanical system is also applicable here. The dissipative and non-dissipative dynamics studied in the optomechanical system and all the plots in Fig. \ref{fig:nb evolution} and \ref{fig:nb dis} will be similar. The optomagnonic parameters that produce similar plots in Fig. \ref{fig:nb evolution} are $G'_{om}=0.5\kappa_h$, $\gamma_m=0.1$ Mhz, $\kappa_h=0.01$ GHz and $n'_{th}=0.5$.
        
    
    \section{Conclusion}
        \label{sec:conc}
        In conclusion, we have studied quantum transduction of superconducting flux-tunable transmon qubit in two hybrid systems: electro-optomechanical and electro-optomagnonical system. The realization and advancement of quantum transduction in such hybrid systems are very crucial for the development of quantum network, quantum internet, etc. The transduction is done in two stages. First, we encode the qubit states in the coherent excitations of mechanical phonon or ferromagnetic sphere magnon without disturbing the qubit state (non-demolition interaction) and in the next stage, we identify these excitations by counting the average number of photon in the optomechanical or optomagnonic WGM cavity. Because of the coherent interaction between the phonon/magnon and the optical photon, the average photon number oscillates with time. The oscillation when the qubit is in the ground state and when in the excited state is exactly opposite. As a result, we can make multiple measurements of the photon number at a regular interval of time and hence know the state of the qubit at each interval. In the presence of dissipation, the optomechanical and optomagnonical coupling strength should be atleast moderately strong in order to perform any measurements before the photon number altogether decays to zero. The required coupling strength in the optomechanical system is extensively studied. But, in the optomagnonic system, the required coupling regime to perform the transduction is not yet explored. However, the possibility of optomagnonic coupling strength going upto $10$ MHz is disscued in \cite{Phys.Rev.Lett.117.123605}. One of the ways to reach such coupling magnitude is to reduce the size of the YIG sphere to few $\mu$m, which is comparable to the size considered in the hybrid system proposed in this work. 
       

    \section*{Acknowledgement}
    RN gratefully acknowledges support of a research fellowship from CSIR, Govt. of India.


    \appendix
    \section{Non-dissipative dynamics.}
        \label{app:dynamics}
        The analytical solution of Eq. \ref{eqn:op_dyna} in the absence of dissipation is given by
        \begin{eqnarray}
            \label{app:eq1}
            \langle\hat{a}^{\dagger}\hat{a}\rangle&=&\frac{1}{2}\{|\alpha_0|^2+|\beta_0|^2+(|\alpha_0|^2-|\beta_0|^2)\,cos(2G_{om}t)\nonumber\\&&-i(\alpha_0^*\beta_0-\alpha_0\beta_0^*)\,sin(2G_{om}t)\}
        \end{eqnarray}
        Here, $|\alpha_0|^2=\langle\hat{a}^{\dagger}\hat{a}\rangle_0$ and $|\beta_0|^2=\langle\hat{b}^{\dagger}\hat{b}\rangle_0$ are the initial values of photon and phonon/magnon. The above equation can be further simplified by simply choosing $|\alpha_0|^2=|\beta_0|^2$. 
        \begin{eqnarray}
            \label{app:eq2}
            \langle\hat{a}^{\dagger}\hat{a}\rangle=|\alpha_0|^2-Im(\alpha_0^*\beta_0)\,sin(2G_{om}t)
        \end{eqnarray}
        The initial coherent amplitudes of phonon/magnon is fixed at $\beta_0=\pm ig_{0}\phi_{ac}\tau$. To keep the oscillatory part in Eq. A2, which is necessary for the transduction, we require that the initial coherent amplitude of the cavity photon $\alpha_0$ should have a non-zero real part. Therefore, we choose $\alpha_0=g_{0}\phi_{ac}\tau$. The oscillation of Eq. A2 
        then becomes
        \begin{eqnarray}
            \label{app:eq3}
            \langle\hat{a}^{\dagger}\hat{a}\rangle=(g_{0}\phi_{ac}\tau)^2\{1-sin(2G_{om}t)\},
        \end{eqnarray}
        when the qubit is in the excited state $(\beta_0=- ig_{0}\phi_{ac}\tau)$, and 
        \begin{eqnarray}
            \label{app:eq4}
            \langle\hat{a}^{\dagger}\hat{a}\rangle=(g_{0}\phi_{ac}\tau)^2\{1+sin(2G_{om}t)\},
        \end{eqnarray}
        when the qubit is in the ground state $(\beta_0= ig_{0}\phi_{ac}\tau)$. 
    \bibliography{reference}

\end{document}